\documentclass{article}

\PassOptionsToPackage{numbers, compress}{natbib}
\usepackage[final]{nips_2016}

\usepackage[utf8]{inputenc} 
\usepackage[T1]{fontenc}    
\usepackage{url}            
\usepackage{booktabs}       
\usepackage{amsfonts}       
\usepackage{nicefrac}       
\usepackage{microtype}      
\usepackage{amsmath}
\usepackage{placeins}
\usepackage{graphicx}
\usepackage{subcaption}
\usepackage{algorithm}
\usepackage[noend]{algpseudocode}
\usepackage{multicol}
\usepackage{multirow}

\title{Multi-stage Clustering of Breast Cancer for Precision Medicine}

\author{
  Chenzhe Qian \\
  \texttt{Columbia University} \\
  \texttt{c.qian@columbia.edu} \\
}

\begin{document}

\maketitle

\begin{abstract}
	Cancer has become one of the most widespread diseases in the world. Specifically, breast cancer is diagnosed more often than any other type of cancer. However, breast cancer patients and their individual tumors are often unique. Identifying the underlying genetic phenotype can lead to precision (personalized) medicine. Tailoring medical treatment strategies to best fit the needs of individual patients can dramatically improve their health. Such an approach requires sufficient knowledge of the patients and the diseases, which is currently unavailable to practitioners. This study focuses on breast cancer and proposes a novel two-stage clustering method to partition patients into hierarchical groups. The first stage is broad grouping, which is based on phenotypes such as demographic information and clinical features. The second stage is fine grouping based on genomic characteristics, such as copy number variation and somatic mutation, of patients in a subgroup resulting from the first stage. Generally, this framework offers a mechanism to mix multiple forms of data, both phenotypic and genomic, to most effectively define individual patients for personalized predictions. This method provides the ability to detect correlation among all factors.
\end{abstract}

\section{Introduction}
Cancer is responsible for one in eight deaths worldwide. It includes more than 100 distinct diseases with diverse risk factors \citep{stratton_cancer_2009}. Breast cancer is one of the most common cancers out there, with more than 1,300,000 cases and 450,000 deaths each year worldwide \citep{koboldt_comprehensive_2012} \citep{ciriello_comprehensive_2015}. It is about twice as common in the first-degree relatives of women with the disease than in the general population, consistent with variation in genetic susceptibility to the disease \citep{easton_genome-wide_2007}.

Recent advances in human genome research opens the door to precision medicine. Precision medicine normally describes the ability to segment heterogeneous subsets of patients whose response to a therapeutic intervention within each subset is homogeneous. In general, it refers to adapting the treatment of a disease to administer the most effective treatment for each person’s disease, get the best results, and avoid unnecessary treatment. Practically, it uses marker-based diagnosis and targeted therapies derived from an individual’s genomic profile.

Most genomic-orientated studies have focused on miRNA sequencing, the frequently used mRNA expression profiling, DNA copy number analysis, single nucleotide polymorphism (SNP) arrays, and whole-exome sequencing \citep{koboldt_comprehensive_2012} \citep{tamborero_comprehensive_2013} \citep{chen_identifying_2013}. On the other hand, a lot of clinical research pays attention to the analysis of  risk factors, pathogenesis characterization, and implications of therapeutic strategies.

However, the roadblocks to converting a genome discovery into a tangible clinical endpoint using the above approaches are numerous and formidable. It is important to establish the biological relevance of a cancer genomic discovery,  realize its clinical potential and discuss some of the major obstacles  \citep{Chin_2011}. Most analyses have used genomic profiles to define broad group distinctions, similar to the use of traditional clinical risk factors. As a result, there remains considerable heterogeneity within the broadly defined groups and predictions are inaccurate for individual patients  \citep{Nevins_2003}.

In this study, we looked at both phenotypes and genotypes to achieve a fine clustering structure of the population, as well as investigate the potential correlation between any pairs of factors. We used Gibbs sampling to approximate the posterior distributions of parameters, and applied mutual information to measure the correlation.

\section{Model and Method}
The model was inspired by the Product Multinomial Mixture Model (PMM) \citep{dunson_2009} and the Simplex Factor Model (SFM) \citep{dunson_2012}. Both are used for modeling high-dimensional unordered categorical data. However, each model has its advantages and limits. PMM seems restrictive since it has a single latent class index for each subject over all variables, while the SFM focuses on a local latent class index for each variable per subject. In other words, the granularity of the PMM is too coarse and the granularity of the SFM is too fine.

This paper proposes a Mixed Factor Model (Figure \ref{mfm}), which has a global latent class index for each subject as well as a local latent class index for some variables. From a biological perspective, the phenotype provides the coarse-grained grouping for the population. For example, the population can be easily distinguished according to gender, age, ethnicity, pathogenesis phase, etc. On the contrary, the genomic characteristics offer a fine-grained information. The diversity of human genome between two individuals is about 0.1\%; furthermore, humans and chimpanzees only have a 1.2\% difference on genome.

\subsection{Mixed Factor Model}
\begin{figure}[h]
	\begin{subfigure}{0.5\textwidth}
		\centering
		\includegraphics[width=0.9\textwidth]{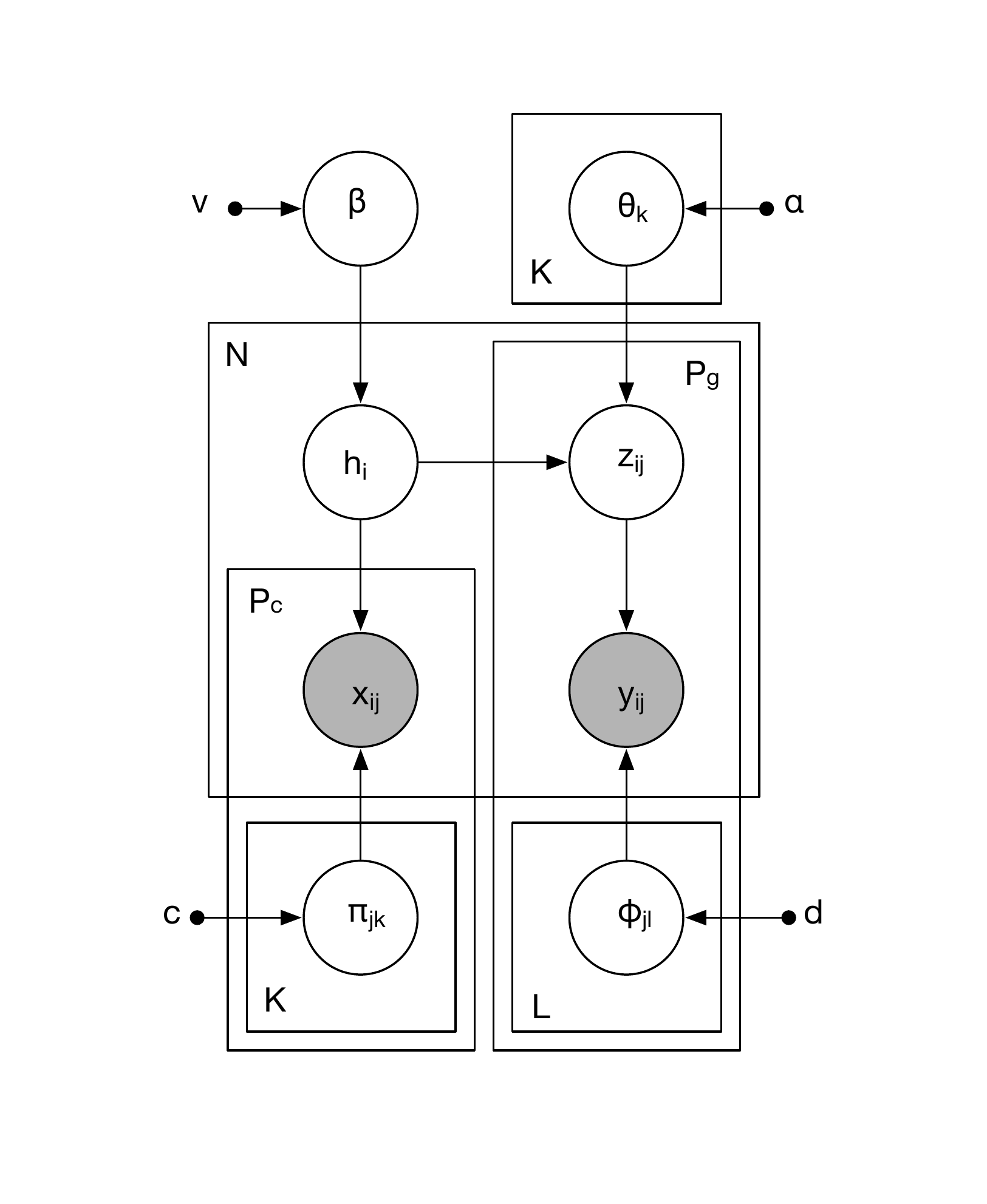}
		\caption{Mixed Factor Model}
		\label{mfm}
	\end{subfigure}
	\begin{subfigure}{0.5\textwidth}
		\centering
		\begin{tabular}{ c | l }
		  \hline
		  $\alpha, \nu, c, d$ & hyperparameters \\ \hline
		  N & The number of patients  \\ \hline
		  $P_c$ & The number of phenotypes \\ \hline
		  $P_g$ & The number of genomic characteristics \\ \hline
		  K & The number of subpopulations \\ \hline
		  L & The number of topics for genomic characteristics \\ \hline
		  $\beta$ & \begin{tabular}{@{}l@{}}Parameters for multinomial distribution for \\ subpopulation\end{tabular} \\ \hline
		  $h_i$ & The indicator of subpopulation \\ \hline
		  $\theta_{k} $ & \begin{tabular}{@{}l@{}}Parameters for multinomial distribution \\ for $k^{th}$ subpopulation\end{tabular} \\ \hline
		  $z_{ij}$ & The indicator of topic assignment for a variable  \\ \hline
		  $\pi_{jk} $ & \begin{tabular}{@{}l@{}}Parameters for multinomial distribution for\\ $j^{th}$ phenotypes of topic k \end{tabular} \\ \hline
		  $\phi_{jl} $ & \begin{tabular}{@{}l@{}}Parameters for multinomial distribution for\\ $j^{th}$ genomic characteristics of topic l \end{tabular} \\ \hline
		  $x_{ij}$ & The observed $j^{th}$ phenotype data for the $i^{th}$ patient\\ \hline
		  $y_{ij}$ & The observed $j^{th}$ genomic data for the $i^{th}$ patient\\ \hline  
		\end{tabular}
		\caption{Summary table of the Model symbols}
	\end{subfigure}
	\caption{Graphic representation of the model and description of its variables}
	\label{model}
\end{figure}

The Mixed Factor Model has a few advantages over both PMM and SFM. It introduces a global latent class to assign an individual to a subgroup, and a local latent class to cluster the variables over one subgroup. More specifically, phenotypes play a key role in clustering the population while the genomic characteristics offer detailed information to support the cluster. Figure \ref{model} describes the model and its variables, and Algorithm \ref{algo} is the generative process of the model. The model is fitted with Gibbs sampling and implemented in C++. For details on the estimation of parameters, please refer to Appendix 6.1

\begin{algorithm}
\caption{Generative process}\label{gp}
	\begin{algorithmic}[]
	\State $\alpha, \nu, c, d$ \Comment{initialize hyperparameters}
	\State $\beta \sim Dir(\nu)$
	\For{each subpopulation $k \in [1..K]$}:
		\State $\theta_{k} \sim Dir(\alpha)$
	\EndFor
	\For{each phenotype $j \in P_c$}:
		\For{each subpopulation $k \in [1..K]$}:
			\State $\pi_{jk} \sim Dir(c)$
		\EndFor
	\EndFor
	\For{each genomic characteristic $j \in P_g$}:
		\For{each topic $l \in [1..L]$}:
			\State $\phi_{jl} \sim Dir(d)$
		\EndFor
	\EndFor
	
	\For{each person $n \in [1..N]$}:
		\State $h_i \sim \mathcal{M}(\beta)$
		\For{each phenotype $j \in P_c$}:
			\State $x_{ij} \sim p(x_{ij} | h_i, \pi_{j}) \sim \mathcal{M}(\pi_{j,h_i})$ \Comment{Generate phenotypes data}
		\EndFor
		\For{each genomic characteristic $j \in P_g$}:
			\State $z_{ij} \sim p(z_{ij} | h_i, \theta) \sim \mathcal{M}(\theta_{h_i})$
			\State $y_{ij} \sim p(y_{ij} | z_{ij}, \phi_{j}) \sim \mathcal{M}(\phi_{j, z_{ij}})$ \Comment{Generate genomic data}
		\EndFor
	\EndFor
	\end{algorithmic}
	\label{algo}
\end{algorithm}
\FloatBarrier

\subsection{Mutual Information}
The second goal of this paper is to draw a potential relationship between any pairs of factors. According to Dunson \citep{dunson_2009}, the measure of association should not depend on the data directly. Rather, it should be a function of the parameters characterizing the multivariate distribution in Bayesian setting. Hence, we can measure the dependence between a pair of variables via mutual information based on the MCMC output. To estimate the correlation between the phenotypes, we consider the pairwise normalized mutual information. 

\begin{align}
	I_{jj'} &= \sum_{c_j=1}^{d_j}\sum_{c_j'=1}^{d_{j'}} \Psi_{c_j c_{j'}}^{jj'} \log (\frac{\Psi_{c_j c_{j'}}^{jj'}}{\Psi_{c_j}^j \Psi_{c_j}^{j'}} ) \label{mi} \\
	H_j &= -\sum_{c_j=1}^{d_j} \Psi_{c_j}^j \log (\Psi_{c_j}^j) \label{entropy} \\
	m_{jj'} &= \frac{I_{jj'}}{\sqrt{H_jH_{j'}}} \label{pmi} \\	
\end{align}

The mutual information $I_{jj'}$ is a general measure of dependence between a pair of random variables, $H_j$ is the marginal entropy of one variable, and $m_{jj'}$ is the normalized mutual information which is akin to the Pearson correlation coefficient.

\subsection{Evaluation}
There are many metrics commonly used to evaluate probabilistic models in scientific literature. In this study, we used predictive likelihood as the metric, which captures the model’s ability to predict a test set of unseen data after having learned its parameters from a training set. We set aside 10\% held-out as a test set and trained on the remaining patients’ data. Then, we computed predictive log likelihood. We followed the approximation of the predictive log likelihood described in Chang’s work \citep{Chang_2009}, in which $p(p^{test}_i|D^{train})$ using $p(p^{test}_i|D^{train}) \approx p(p^{test}_i|\hat\theta)$, where $\hat\theta$ representing point estimation of hidden variables from posterior distributions. We used PMM and SFM as base models to compare the performance of the proposed model.

\section{Results and Discussion}
The experiments were conducted over the real data containing 772 out of 1,097 patients. Patients who had missing variables were not considered. The number of the subgroup and the number of the latent class were both determined by grid search. The optimal number of the subgroup was 5 and the optimal number of the latent class was 4 in terms of the log likelihood. 

\begin{figure}[h]
	\begin{subfigure}{0.5\textwidth}
		\centering
		\begin{tabular}{ c | c | c }
		  \hline
		  \# Group & mean & std \\		
		  \hline
			0 & 60 & 0.8 \\
			1 & 80 & 1.2 \\
			2 & 40 & 0.7\\
			3 & 50 & 0.5 \\
			4 & 55 & 0.4\\
		  \hline  
		\end{tabular}
		\caption{The mean age with standard deviation of each group}
		\label{age}
	\end{subfigure}
	\begin{subfigure}{0.5\textwidth}
		\centering
		\includegraphics[width=0.7\textwidth]{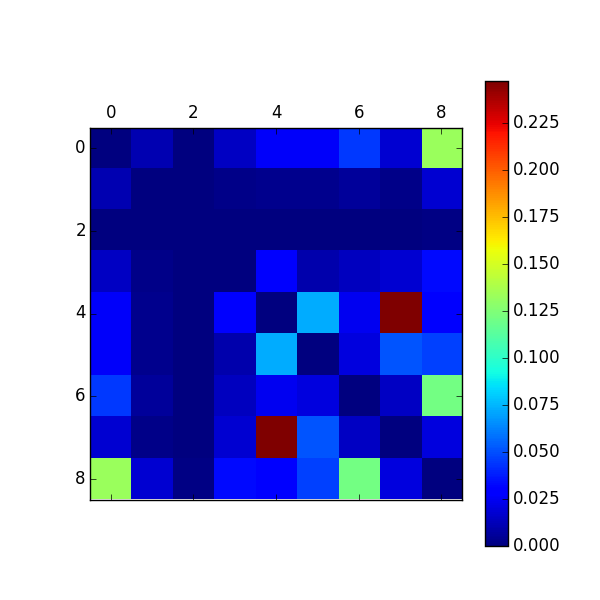}
		\caption{The correlation matrix of phenotypes}
		\label{corrMat}
		\end{subfigure}
	\caption{Interpretation of model estimation}
	\label{postDist}
\end{figure}

The interpretation of the fitted model was reasonable. Each subgroup successfully captured its own feature. For example, Figure \ref{age} showed that each group had its own peaked age range, and the peaked range was distinct among all groups except for group 2. More importantly, the correlation measured by the model was also plausible. Figure \ref{corrMat} shows that the pathologic stage (\#8) was strongly associated with the pathologic M (Distant Metastasis) (\#0). This relationship could not be captured in a pairwise Poisson regression. Similarly, the correlation between lymph node count (\#5) and age (\#4) was also detected. The most intense correlation between \#4 and \#7 was age and ethnicity. Many studies have proved that race has a significant impact on the age of cancer effects. This result further supports the model.

We measured the predictive log likelihood of PMM, SFM and the proposed mixed factor model with same training and testing data. The mixed factor model held highest predictive log likelihood of -137, whereas PMM and SFM scored at -320 and -171 respectively. Clearly, the proposed model had best performance. We found that the performance of PMM dramatically declined if genomic data added. As the genomic data was granular and highly specific, for PMM, adding this might be equivalent to adding noise. We also saw that, in SFM, the number of optimal clusters for phenotypes tended to be smaller than that for genomic data. Though it was expected, SFM was less flexible than the proposed model. This probably explained the performance disparity between SFM and the proposed model.

In summary, the key idea of the model is to split the observed data using two different levels of granularity. The coarse-grained variables are the dominant factors for clustering a set of subjects, while the fine-grained variables will compensate with detailed information to improve the accuracy. Another feature of this model is that, without assuming the dependence structure of the variables, one can infer the correlation among the variables.

\FloatBarrier

\FloatBarrier


\section*{Acknowledgments}
I would like to thank David Blei, Rajesh Ranganath and Da Tang for their valuable suggestions and advice.

\bibliographystyle{unsrt}
\bibliography{ref}


\section{Appendix}
\subsection{Inference and Parameter Estimation}
We used the common Gibbs sampling to draw samples from the posterior distributions of all parameters for the hierarchical model defined in Alg. \ref{gp}. The default hyperparameters specification can be modified if one has prior information of the category probabilities and/or the number of subpopulations. 

\textbf{Step 1} For z = 1,..,L, update $\phi_{zj}$ from the Dirichlet full conditional posterior distribution
\begin{align}
	p(\phi_{jl} | y_{ij}, z_{ij}) = Dir(d+\sum_i^N z_{ij}^l y_{ij})
\end{align}

\textbf{Step 2} For h = 1,..,K, update $\pi_{hj}$ from the Dirichlet full conditional posterior distribution
\begin{align}
	p(\pi_{jk} | x_{ij}, h_i) = Dir(c+\sum_i^N h_i^k x_{ij}) 
\end{align}

\textbf{Step 3} For i=1,..,N, For j=1..P, update $z_{ij}$ by sampling from the multinomial full conditional posterior distribution 
\begin{align}
	p(z_{ij} = l | \cdot ) &= \frac{ \exp(\log \Gamma + y_{ij}^T \log\phi_{j,l})\theta_{h_i,l} } {\sum_{l'}^L \exp(\log \Gamma + y_{ij}^T \log\phi_{j,l'}) \theta_{h_i,l'}}
\end{align}

\textbf{Step 4} For i=1,..,N, update $h_i$ by sampling from the multinomial full conditional posterior distribution 
\begin{align}
	p(h_i = k | \cdot ) &= \frac{\beta_k \prod_j^{Pc}\pi_{j, k, x_{ij}} \prod_j^{Pg} \theta_{k,z_{ij}} } {\sum_{k'}^K (\beta_{k'} \prod_j^{Pc}\pi_{j, k', x_{ij}} \prod_j^{Pg} \theta_{k',z_{ij}} ) }
\end{align}

\textbf{Step 5} For h = 1,..,K, update $\theta_k$ from the Dirichlet full conditional posterior distribution
\begin{align}
	p(\theta_{k} | z_{ij}, h_i) = Dir(\alpha + \sum_i^N h_i^k \sum_j^{Pg} z_{ij})
\end{align}

\textbf{Step 6} Update $\beta$ from the Dirichlet full conditional posterior distribution
\begin{align}
	p(\beta |\nu, \pmb{h}) = Dir(\nu + \sum_i^{N} h_i^k)
\end{align}

\subsection{Evaluation}
The predictive likelihood per patient is defined as:
\begin{align*}
	p(x_i, y_i | \bf{X}^{train}, \bf{Y}^{train}) &= \int_{\hat{\pi}} \int_{\hat{\phi}} p(x_i, y_i | \hat{\pi},  \hat{\phi}) p(\hat{\pi},  \hat{\phi}|\bf{X}^{train}, \bf{Y}^{train}) d \hat\pi d\hat\phi \\
	& \text{  where, } \hat{\pi} \text{ and } \hat\phi \text{ is point estimate of posterior distribution} \nonumber \\
\end{align*}

\end{document}